

\input harvmac

\def\np#1#2#3{Nucl. Phys. B {#1} (#2) #3}
\def\pl#1#2#3{Phys. Lett. B {#1} (#2) #3}

\def\prl#1#2#3{Phys. Rev. Lett. {#1} (#2) #3}
\def\physrev#1#2#3{Phys. Rev. D {#1} (#2) #3}
\def\prd#1#2#3{Phys. Rev. D {#1} (#2) #3}

\def\prep#1#2#3{Phys. Rep. {#1} (#2) #3}

\def\pf{{\rm Pf ~}}
\def\pfp{{\rm Pf}^{\prime}~}

\baselineskip=12pt
\hfill{SLAC-PUB-95-7041} 

\hfill{IASSNS-HEP-95/114} 

\hfill{NSF-ITP-95-149}

\hfill{hep-th/9603158}

\baselineskip=12pt
\Title{} 
{\vbox{\centerline{Dynamical Supersymmetry Breaking} 
\vglue .1in
\centerline{on Quantum Moduli Spaces}}}

\bigskip
\centerline{\tenbf Kenneth Intriligator}
\vglue .1cm
\centerline{Institute for Advanced Study}
\centerline{Princeton, NJ 08540}

\vglue .2in 

\centerline{\tenbf Scott Thomas}
\vglue .1cm
\centerline{Stanford Linear Accelerator Center}
\centerline{Stanford University}
\centerline{Stanford, CA 94309}
\centerline{and}
\centerline{Institute for Theoretical Physics}
\centerline{University of California}
\centerline{Santa Barbara, CA  93106}

\bigskip

\noindent

Supersymmetry breaking by the quantum deformation 
of a classical moduli space is considered. 
A simple, non-chiral, renormalizable model is presented to 
illustrate this mechanism. 
The well known, chiral, $SU(3) \times SU(2)$ 
model and its generalizations are shown to 
break supersymmetry by this mechanism
in the limit $\Lambda_2 \gg \Lambda_3$.  
Other supersymmetry breaking models, with classical flat directions 
that are only lifted quantum mechanically, are presented. 
Finally, by integrating in vector matter, the strongly coupled region
of chiral models with a dynamically generated
superpotential is shown to be continuously connected to a weakly
coupled description in terms of confined degrees of freedom,
with supersymmetry broken at tree level.

\Date{3/96}

\lref\intpou{K. Intriligator and P. Pouliot, hep-th/9505006, 
\pl{353}{1995}{471}.}

\lref\suf{I. Affleck, M. Dine, and N. Seiberg, \pl{137}{1984}{187};
\prl{52}{1984}{1677}.}

\lref\ven{Y. Meurice and G. Veneziano, \pl{141}{1984}{69}.}

\lref\ads{I. Affleck, M. Dine, and N. Seiberg, \np{241}{1984}{493};
\np{256}{1985}{557}.}%

\lref\amati{D. Amati, K. Konishi, Y. Meurice, G. Rossi, and 
G. Veneziano, \prep{162}{1988}{169}.}

\lref\ils{K. Intriligator, R.G. Leigh and N. Seiberg, hep-th/9403198,
\physrev{50}{1994}{1092}.}%

\lref\sem{N. Seiberg, hep-th/9411149, \np{435}{1995}{129}.}%

\lref\iss{K. Intriligator, N. Seiberg, and S. Shenker,
hep-ph/9410203,
\pl{342}{1995}{152}.}

\lref\brp{J. Bagger, L. Randall, and E. Poppitz, hep-ph/9405345,
\np{426}{1994}{3}.}

\lref\nati{N. Seiberg, hep-th/9402044, \physrev{49}{1994}{6857}.}%

\lref\nonren{N. Seiberg, hep-ph/9309335, \pl{318}{1993}{469}.}

\lref\swi{N. Seiberg and E. Witten, hep-th/9407087,
\np{426}{1994}{19}.}%
 
\lref\sn{A. Nelson and N. Seiberg, hep-ph/9309299, \np{426}{1994}{46}.}

\lref\hitoshi{H. Murayama, hep-ph/9505082, \pl{355}{1995}{187}.}

\lref\pt{E. Poppitz and S. Trivedi, hep-th/9507169,
\pl{365}{125}{1996}.}

\lref\panti{P. Pouliot, hep-th/9510148, \np{367}{151}{1996}.}

\lref\bkn{M. Dine and D. Macintire, hep-ph/9205227, 
\physrev{46}{2594}{1992};
T. Banks, D. Kaplan, and A. Nelson, hep-ph/9308292,
\physrev{49}{1994}{779}.}

\lref\smallmu{G. Farrar, RU-95-25, hep-ph/9508291;
J. Feng, N. Polonsky, and S. Thomas,
hep-ph/9511324, \pl{370}{95}{1996}.}

\lref\tools{M. Dine, A. Nelson, Y. Nir, and Y. Shirman,
hep-ph/9507378, \prd{53}{1996}{2658}.}

\lref\visible{M. Dine and A. Nelson, hep-ph/9303230, 
\physrev{48}{1993}{1277};
M. Dine, A. Nelson, and Y. Shirman, hep-ph/9408384,
\physrev{51}{1995}{1362}.}

\lref\index{E. Witten, \np{202}{1982}{253}.} 

\lref\poustr{P. Pouliot and M.J. Strassler, RU-95-78, hep-th/9602031.}

\lref\talk{
K. Intriligator and S. Thomas, talk presented at {\it Unification:
{}From the Weak Scale to Planck Scale}, Institute for Theoretical
Physics, Santa Barbara, CA, Oct. 23, 1995.}

\lref\toappear{K. Intriligator and S. Thomas, to appear}

\baselineskip=16pt

\newsec{Introduction}

Non-perturbative gauge dynamics can lead to supersymmetry breaking in
theories in which supersymmetry is unbroken at tree level.  The
``classic'' models of dynamical supersymmetry breaking
\refs{\suf, \ads} all
relied on dynamically generated superpotentials.  This leads to a
potential which lifts the classical moduli space and drives scalar
fields to large expectation values.  Additional tree level
interactions can give a potential which rises at large expectation
values, thereby leading to a stable ground state.  In certain
circumstances the auxiliary components and potential do not vanish in
such a ground state, and supersymmetry is spontaneously broken.
Recent advances led by Seiberg \refs{\nati, \sem}, however, have shown
that many supersymmetric theories have other types of non-perturbative
dynamics, which lead to exactly degenerate quantum moduli spaces of
vacua rather than dynamically generated superpotentials.  In some
cases the quantum moduli space is smooth and the low energy theory
consists only of the massless moduli everywhere on the space.  In
other cases, there are additional non-perturbative light fields at
various points on the moduli space. See
\ref\isrev{K. Intriligator and N.  Seiberg, hep-th/9509066} for a
recent review and references.  The quantum dynamics of theories with
moduli spaces of vacua could in principle lead to new mechanisms for
supersymmetry breaking, even though the dynamical superpotential
exactly vanishes on the classical moduli space. 
The possibility of breaking supersymmetry as the
result of a smooth quantum moduli space (with confinement at the
origin) was considered in a simple model in Ref.
\iss.  Here we consider models in which
supersymmetry is broken as a result of the quantum deformation of a
classical moduli space constraint.

The mechanism of supersymmetry breaking by the quantum deformation of
a moduli space relies on the observation that the patterns of
breakings for global or gauge symmetries on a quantum moduli space may
differ from those on the classical moduli space.  For example, since
moduli typically transform under global symmetries, there is a point
on the classical moduli space at which all the fields have zero
expectation value, and the global symmetries are unbroken.  However,
in the quantum theory some of the global symmetries can remain broken
everywhere on the moduli space \nati.  Points which are part of the
classical moduli space can therefore be removed by the quantum
deformation.  If tree level interactions give vanishing potential and
auxiliary components only at points on the classical moduli space
which are not part of the quantum deformed moduli space, supersymmetry
is broken in the quantum theory.

The superpotential and spectrum of light fields on moduli spaces of
vacua can often be determined exactly.  However, in theories in which
supersymmetry is broken, a quantitative discussion of the ground state
and excited spectrum also requires knowledge of the Kahler potential.
This can only be approximately determined in certain limits.  One such
limit is when the expectation values in the ground state are much
larger than all dynamical scales.  The gauge dynamics are then weak,
and the Kahler potential is approximately the classical one for the
underlying fields projected onto the classical moduli space.  As we
illustrate in several examples, another limit in which the Kahler
potential can be approximately determined is near points of enhanced
symmetry on the quantum moduli space of vacua.  In this limit it is
often possible to use 't Hooft anomaly matching conditions to find the
correct degrees of freedom.  For small expectation values away from
such points the Kahler potential should be approximately canonical in
these degrees of freedom, up to small corrections suppressed by the
dynamical scale.  This information about the Kahler potential played
an important role in
\iss\ and will be exploited here as well.  It allows an analysis in
regions which may have naively been thought to be strongly coupled and
inaccessible.

In the next section we present the simplest example of supersymmetry
breaking via the quantum deformation of a classical moduli space
constraint.  Taking the Yukawa coupling in the model to be small
allows an analysis in various limits by perturbing about the quantum
moduli space.  This model also provides an example of a renormalizable
model in which singlet fields participate directly in the
supersymmetry breaking.  In the third section, we discuss the well
known $SU(3)\times SU(2)$ model of Affleck, Dine, and Seiberg \ads.
We show that supersymmetry breaking in the limit $\Lambda _2\gg\Lambda
_3$ is associated with the quantum deformation of a classical moduli
space constraint due to the $SU(2)$ dynamics, rather than a
dynamically generated superpotential arising from the $SU(3)$
dynamics.  In the fourth section we present some generalizations of
this model and determine the scaling of the vacuum energy in various
limits.  In the fifth section, we present models which break
supersymmetry even though they have classical flat directions which
are not lifted by the tree level superpotential.  These directions are
only lifted by the tree level superpotential in the presence of
quantum effects.  In section six we discuss the addition of
vector-like matter to theories which break supersymmetry by a
dynamically generated superpotential.  Taking the mass of the
vector-like matter much less than the dynamical scales allows these
theories to be connected to a weakly coupled description in terms of
confined degrees of freedom.  Demonstrating supersymmetry breaking in
the theory with extra matter gives confidence that in the original
theory there are not additional branches at strong coupling along
which supersymmetry is restored.  These models also provide examples
of supersymmetry breaking based on confinement.  Finally, in section
seven we present our conclusions and discuss possible applications of
the simple model of section two.  We also comment on general features
of non-chiral models which break supersymmetry, and point out how the
Witten index can vanish in such models.

\newsec{$SU(2)$ Quantum Moduli Space with Singlets}

The simplest example of a moduli space with a quantum deformed
constraint is $SU(2)$ with four doublet matter fields $Q_i$,
$i=1,\dots,4$.  The classical moduli space is parameterized by the
gauge invariants $M_{ij}= Q_i Q_j$ subject to the constraint $ {\rm
Pf}~M \equiv \epsilon^{ijkl} M_{ij} M_{kl} = 0$.  Quantum
mechanically, the constraint is modified to $ {\rm Pf}~M =
\Lambda_2^4$ \nati.  While the point $M_{ij}=0$ is part of the
classical moduli space, it does not lie on the quantum moduli space.

Supersymmetry would be broken if the quantum modification of a moduli
space were incompatible with a stationary superpotential, $W_{\phi}
\neq 0$, where $\phi$ is any field.  For example, in the $SU(2)$ case
mentioned above, supersymmetry would be broken if there were $F$ terms
which only vanished for $M_{ij}=0$.  A simple realization of this is
to add to the $SU(2)$ model given above, six singlet fields
$S^{ij}=-S^{ji}$, where $i,j=1,\dots,4$, with couplings \talk\
\eqn\sqq{
W_{tree} = \lambda ~ S^{ij} Q_i Q_j = \lambda~S^{ij} M_{ij}.  } This
superpotential leaves invariant an $SU(4)_F$ flavor symmetry under
which $Q_i$ transform as $\bf{4}$, and $S^{ij}$ transform as $\bf{6}$.
There is also an anomaly free $U(1)_R$ symmetry under which $R(Q)=0$
and $R(S)=2$.  Any mass terms $mQQ$ can be absorbed in a shift of the
$S^{ij}$.  The form of the couplings \sqq\ can be enforced by discrete
symmetries, or by weakly gauging certain subgroups of the flavor
symmetry.  The absence of any higher order terms in $S$ which respect
the $SU(4)_F$ flavor symmetry, such as $({\rm Pf}~S)^k$, can be
enforced by gauging a discrete subgroup of $U(1)_R$.  Because the
fields $S^{ij}$ do not appear alone to any power in the
superpotential, they are similar to the moduli of string theory which
parameterize flat directions of the perturbative potential.
Classically, there is a moduli space of supersymmetric vacua with
$M_{ij}=0$ and $S^{ij}$ arbitrary.  Quantum mechanically, the $S^{ij}$
equations of motion, $\lambda M_{ij}=0$, are incompatible with the
quantum constraint ${\rm Pf}~M=\Lambda_2^4$.  The classical moduli
space of supersymmetric vacua is completely lifted for $\lambda \neq
0$ as a result
of the quantum modification of the $SU(2)$ moduli space,
and supersymmetry is broken. 
 
The modification of the moduli space can be realized in the
superpotential with a Lagrange multiplier to enforce the quantum
constraint.  The full superpotential can then be written as
\eqn\sutexact{
W = \lambda~S M + {\cal A} ( {\rm Pf}~M - \Lambda_2^4), } where ${\cal
A}$ is a Lagrange multiplier field.  For $\lambda \ll 1$ the
supersymmetry breaking ground state lies close in field space to the
$SU(2)$ quantum moduli space, so the model may be analyzed by
perturbing about this space.  This amounts to enforcing the ${\cal A}$
equation of motion and analyzing the resulting potential in terms of
the remaining light degrees of freedom.  This is physically
reasonable, as the modes which take the system away from the quantum
moduli space are expected to have mass of at least ${\cal
O}(\Lambda_2)$.  For $\lambda \neq 0$ the heavy and light states of
course mix, as shown below explicitly.  However, for $\lambda \ll 1$
the mixing is small and does not affect the leading behavior of
physical quantities such as the vacuum energy.  For $\lambda \sim 1$
supersymmetry is still broken but can not be described in any
quantitative way in terms of the $S^{ij}$ and $M_{ij}$ degrees of
freedom.

The description of supersymmetry breaking in the $\lambda \ll 1$ limit
depends on the values $\lambda S^{ij}$.  For $\lambda S^{ij} \ll
\Lambda_2$, the mesons $M_{ij}$ are light and should not be integrated
out.  For $\lambda =0$, the quantum constraint breaks the $SU(4)_F$
flavor symmetry at a generic point to $SU(2)_F \times SU(2)_F$, and at
an enhanced symmetry point to $SP(2)_F \sim SO(5)_F$ \nati.  Enhanced
symmetry points are always extrema of the full potential, so are good
candidates for minima.  At the enhanced symmetry point the $M_{ij}$
have the form
\eqn\Mzero{  
M_0 = \pm \left(
\matrix{  
 i \sigma_2 & \cr & i \sigma_2 \cr } \right) \Lambda _2^2.}  The
fluctuations of $M^{ij}$ away from \Mzero\ subject to the quantum
constraint $\pf M=\Lambda _2^4$ are fields $M_5$ in the $\bf{5}$ of
$SO(5)_F$.  The fields $\widehat{M}_5 = M_5 / \Lambda_2$ comprise the
massless spectrum near the vacuum \Mzero, as evidenced by the 't Hooft
anomaly matching for the unbroken flavor symmetry \nati.  Because
these fields are the relevant degrees of freedom at
\Mzero, for small expectation values they have canonical Kahler
potential up to small corrections, $K = \widehat{M}_5^{\dagger}
\widehat{M}_5 g(t=\widehat{M}_5^{\dagger} \widehat{M}_5 /
|\Lambda_2|^2)$ where $g(0)=1$.  Under the $SO(5)_F$ the fields $S^{ij}$
break up as $S_5 \oplus S_0$ transforming as $\bf{5}$ and $\bf{1}$
respectively.
At the enhanced symmetry point, 
the $\lambda SM$ term in the superpotential yields  
\eqn\snot{
W = \lambda M_5 S_5 \pm 2 \lambda \Lambda_2^2 S_0.  } The first term
mixes the $\widehat{M}_5$ and $S_5$ moduli which are present for
$\lambda =0$.  The resulting Dirac states receive a small mass $m =
\lambda \Lambda_2$.  The second term is linear in the singlet
component $S_0$.  Supersymmetry is therefore broken by the $F$
component of $S_0$ with two degenerate vacua of energy $V \sim
|\lambda^2 \Lambda _2^4|$.

For $\lambda S^{ij} \gg \Lambda_2$, the fields $Q_i$ are heavy and can
be integrated out by enforcing the $M_{ij}$ equations of motion.
Treating $\lambda$ as a small parameter and restricting to the
$M_{ij}$ quantum moduli space, the superpotential
\sutexact\ in the $\lambda S^{ij} \gg \Lambda_2$ limit becomes
\eqn\supf{
W = \pm 2 \lambda \Lambda_2^2 \sqrt{{\rm Pf}~S}.  } Because $\pf
({\partial W / \partial S})=\lambda ^2\Lambda _2^4\neq 0$,
supersymmetry is broken.  In terms of the $S^{ij}$ components $S_0$
and $S_5$ introduced above, $\sqrt{ \pf S}=\sqrt{S_0^2-S_5^2}\approx
S_0$ for $S_0^2\gg S_5^2$.  The superpotential \supf\ amounts to a
linear term for $S_0$ and gives $S_5$ a mass of order $m \sim \lambda
\Lambda_2^2 / S_0$.  In this limit supersymmetry is again broken by
the $F$ component of $S_0$ with two vacua of energy $V \sim |\lambda^2
\Lambda_2^4|$.

The quantum modification of the moduli space, which breaks
supersymmetry, can be given another interpretation in the $\lambda
S^{ij} \gg \Lambda_2$ limit.  In this limit, the quarks $Q$ are
massive and can be integrated out as described above.  The low energy
theory then consists of the singlets $S^{ij}$ along with pure $SU(2)$
Yang-Mills theory with a scale $\Lambda_L$ related to that of the high
energy theory by the matching condition $\Lambda_L^6 = \lambda ^2(\pf
S) \Lambda_2^4$.  Gaugino condensation in the pure $SU(2)$ theory
results in a superpotential $W_L = \pm 2 \Lambda_L^3$, giving
precisely the superpotential \supf.  So supersymmetry breaking is
described in terms of a quantum deformation of the classical moduli
space for $\lambda S^{ij} \ll \Lambda_2$ while, for $\lambda S^{ij}
\gg \Lambda_2$, it is possible to describe it in terms of gaugino
condensation.

Note that the description of supersymmetry breaking in terms of
gaugino condensation is entirely at the renormalizable level.  This is
in contrast to the usual discussion of supersymmetry breaking by
gaugino condensation (within supergravity for example) which requires
a non-renormalizable, moduli dependent gauge kinetic function.  In
addition, the usual discussion requires another mass scale to define a
nontrivial curvature on the Kahler manifold of the modulus (or
equivalently a non-linear gauge kinetic function) in order to
stabilize the gaugino condensate against the Dine-Seiberg instability
\ref\ds{M.  Dine and N. Seiberg, \pl{162}{299}{1985}}.  Without this,
the theory is driven to weak coupling and supersymmetry is restored.
Here, however, no additional mass scale is required to obtain stable
supersymmetry breaking by gaugino condensation.

This model has a pseudo-flat direction corresponding to the $S_0$
component of $S^{ij}$ along which $V \sim |\lambda^2 \Lambda_2^4|$.
This direction would be exactly flat if the Kahler potential for $S$
were precisely canonical.  Quantum contributions to the Kahler
potential $K=S_0^\dagger S_0f(t=|\Lambda _2|^2/S_0^\dagger S_0)$ can
lift this degeneracy, giving the scalar potential $V(t) = 4 |\lambda
^2\Lambda _2^4|(f-tf'+t^2f'')^{-1}$.  The dominant quantum
correction to the potential comes from integrating out the lightest
states which, in the limit $\lambda S_0 \gg \Lambda _2$, are the fields
$S_5$ with mass $m\sim \lambda \Lambda _2^2/S_0\ll \Lambda _2$.
However, because these fields receive a supersymmetric mass to lowest
order, this contribution to the vacuum energy vanishes at this order.
The Kahler potential for $S_0$ is then smooth as $S_0 \rightarrow
\infty$ and $f(0)=1$.  For finite $S_0$ there are corrections to the
Kahler potential.  Unfortunately, because the strongly coupled excited
spectrum is incalculable, the precise function $f(t)$ can not be
obtained.  The natural scales for the minima along this direction are
either $S_0 = 0, {\cal O}(\Lambda_2),$ or $\infty$.  If the minimum
does lie at $S^{ij}=0$, the $U(1)_R$ symmetry remains unbroken;
otherwise, there is an $R$-axion in the massless spectrum.  As
emphasized in \refs{\nonren,\sn} 
and also seen in \iss, this illustrates that a
spontaneously broken $U(1)_R$ symmetry is a sufficient but not
necessary condition for supersymmetry breaking.  If the minima are at
$S_0 \rightarrow \infty$, it may appear formally that the theory does
not have a stable ground state.  This is of course irrelevant to the
question of supersymmetry breaking since the potential approaches a
non-zero constant as $S_0 \rightarrow \infty$.

The existence of this pseudo-flat direction may seem to contradict the
common lore that models of dynamical supersymmetry breaking can not
have flat directions.  In contrast to flat directions along which a
gauge group is Higgsed and becomes weaker, here the matter fields
become more massive and the theory becomes more strongly coupled,
leading to a vacuum energy which does not vanish even infinitely far
along the pseudo-flat direction.  

The scheme of obtaining supersymmetry breaking as the result of the
quantum deformation of a moduli space by coupling singlets (or fields
with weak gauge charges) to meson bilinears is easily extended to
other gauge groups, such as $SP(N_c)$ with $N_f=N_c+1$ flavors of
fundamental quarks \intpou.  Another generalization is to groups in
which higher order invariants are required to parameterize the quantum
moduli space, such as $SU(N_c)$ with $N_f=N_c$ flavors \nati.  In this
case, singlets must couple to each invariant which appears in the
quantum constraint; otherwise, supersymmetry would be restored on some
subspace.  For example, in $SU(N_c)$ with $N_c$ flavors of fundamental
quarks $Q_i$ and $\overline{Q}_j$, two singlets, $S$ and
$\overline{S}$, may be coupled to the baryon invariants $B=Q^{N_c}$
and $\overline{B}=\overline{Q}^{N_c}$ as $W=\lambda_1 \overline{S}B +
\lambda_2 S \overline{B}$.  This coupling may be enforced, for
example, by gauging baryon number.  For $\lambda_i \neq 0$ this
coupling lifts the $B \overline{B} \neq 0$ branch of the moduli space.
However, the quantum constraint ${\rm det}~M - B\overline{B} =
\Lambda^{2N_c}$
\nati, where $M_{ij}=Q_i \overline{Q}_j$, 
is still satisfied on a moduli space parameterized
by the $M_{ij}$ subject to 
${\rm det}~M = \Lambda^{2N_c}$, with 
$S=\overline{S}=B=\overline{B}=0$.
Coupling additional fields to all the $M_{ij}$ would 
completely lift the moduli space of supersymmetric
vacua and lead to supersymmetry breaking.

\newsec{The $SU(3)\times SU(2)$ Model}

Perhaps the best known model of dynamical supersymmetry breaking is
the $SU(3)\times SU(2)$ model of Affleck, Dine, and Seiberg \ads.  The
usual discussion of supersymmetry breaking in this model has
implicitly focused on the limit where the $SU(3)$ dynamics dominates
and supersymmetry breaking is associated with a dynamically generated
superpotential.  In this section, we show that, in a limit where the
$SU(2)$ dynamics dominates, supersymmetry breaking is associated with
a quantum-deformed moduli space constraint.

The matter content of the model is 
\eqn\ttfields{
\matrix{  
\quad   & SU(3) \times SU(2) \cr
        &             \cr
P       & (3,2)       \cr
L       & (1,2)       \cr
\overline{U} & (\overline{3},1) \cr
\overline{D} & (\overline{3},1). \cr 
}} This is just the one generation supersymmetric standard model
without hypercharge, the positron, or Higgs bosons.  Classically, this
model has a moduli space parameterized by three invariants: $Z=P^2
\overline U \overline D$, $X_1 = P L \overline D$, and $X_2 = P L
\overline U$.  There is another gauge invariant, $Y=P^3 L$, which
vanishes classically by Bose statistics of the underlying fields.  The
gauge group is completely broken for generic vacua on the classical
moduli space; the above invariants are the fields which are left
massless after the Higgs mechanism.  At tree level there is a single
renormalizable coupling which can be added to the superpotential,
\eqn\tttree{
W_{tree} = \lambda X_1.  } This superpotential leaves invariant
non-anomalous accidental 
$U(1)_R$ and $U(1)$ flavor symmetries, and 
completely lifts the classical moduli space.  Classically,
there is a supersymmetric ground state at the origin,
with the gauge symmetries unbroken. 

Non-perturbative gauge dynamics generate an additional term in the
effective superpotential; the exact effective superpotential
is fixed by holomorphy, symmetries, and an instanton
calculation to be 
\eqn\wsuiii{
W={\Lambda _3^7\over Z}+
{\cal A}(Y- \Lambda_2^4) +
\lambda X_1.  
}
where ${\cal A}$ is a Lagrange multiplier field. 
The first term is generated by an instanton in the broken
$SU(3)$;
it is just the dynamical superpotential which would arise
over the classical moduli space for $\Lambda_2=0$.
The second term enforces the quantum deformed constraint 
$Y=P^3L=\Lambda_2^4$.
This constraint can be seen in the limit 
$\Lambda_2 \gg \Lambda_3$. 
In this limit the $SU(3)$ is weakly gauged at the scale $\Lambda_2$.
The $SU(2)$ theory therefore has two flavors with a quantum deformed
constraint \nati.
Assuming the full Kahler potential is positive definite and nonsingular,
\wsuiii\ lifts the classical ground state
$Z=X_i=0$.  In the ground state of the quantum theory, both 
the $U(1)_R$ and supersymmetry 
are spontaneously broken \refs{\ads, \nonren, \sn}.  

The description of supersymmetry breaking depends on the relative
importance of the dynamically generated $SU(3)$ superpotential and the
$SU(2)$ deformation of the moduli space.  In the limit $\Lambda_3 \gg
\Lambda_2$ the $SU(2)$ is weakly gauged at the scale $\Lambda_3$.  The
theory can then be analyzed by perturbing about the classical moduli
space, $Y=0$, ignoring the second term in \wsuiii.  For $\lambda \ll
1$ the vacuum expectation values of the fields in the ground state are
close to the classical moduli space, and large compared to both
$\Lambda_2$ and $\Lambda_3$.  Both $SU(3)$ and $SU(2)$ are then
Higgsed at a high scale and thus weakly coupled.  In this weak
coupling limit the Kahler potential $K$ may be approximated by the
classical Kahler potential $K_{cl}$ for the elementary fields
projected on the classical moduli space \ads.  The vacuum expectation
values of the fields and ground state energy can then be computed
numerically in this limit \refs{\ads, \brp}.  Parametrically, for
$\lambda \ll 1$, the field expectation values and vacuum energy scale
as $\phi \sim \Lambda_3 / \lambda^{1/7}$ and $V \sim |\lambda^2 (
\Lambda_3 / \lambda^{1/7})^4| = |\lambda^{10/7} \Lambda_3^4|$.  In
order for this approximation to be valid, the quantum deformation of
the moduli space must be unimportant at the scale of the expectation
values, which requires $\Lambda_3 \gg \lambda^{1/7} \Lambda_2$.
Notice that in this limit the $SU(2)$ acts a spectator in the
non-perturbative dynamics which break supersymmetry.  It restricts
certain couplings which would otherwise be allowed in the
superpotential, and it's classical gauge potential lifts certain
directions in field space.

For $\Lambda_2 \gg \Lambda_3$ the $SU(3)$ is weakly gauged at the
scale $\Lambda_2$ and the quantum modification of the moduli space
arising from the second term in \wsuiii\ is important.  For $\lambda
\ll 1$, the relevant moduli space to perturb about is then the quantum
one.  Below the scale $\Lambda_2$, the theory can be described in
terms of the light $SU(2)$ singlet fields $\widehat{q} = PL/
\Lambda_2$ in the $\bf 3$ of $SU(3)$ and $\widehat{\overline q} = P^2/
\Lambda_2$, $\overline U$, and $\overline D$ all in the $\overline
{\bf 3}$ of $SU(3)$, subject to the quantum constraint $\widehat{q}
\widehat{\overline q} = \Lambda_2^2$.  The components of $\widehat{q}$
and $\widehat{\overline q}$ which preserve the constraint are just the
components $\widehat{M}_5$ discussed in the previous section.  On the
quantum moduli space the $SU(3)$ is generically completely broken.  At
the point ${\overline U} = {\overline D} =0$ and $\widehat{q} =
\widehat{\overline q} = \Lambda_2$ there is an unbroken
$SU(2)^{\prime} \subset SU(3)$.  Unlike the classical moduli space,
there is no point at which the $SU(3)$ is restored.\foot{Note that the
fields $\widehat{q}$, $\widehat{\overline q}$, $\overline U$, and
$\overline D$ do not form an anomaly free representation of $SU(3)$.
This is not inconsistent since $SU(3)$ is never restored on the
quantum moduli space.}  For $\lambda \neq 0$, the classical
supersymmetric ground state at which the full gauge symmetry is
unbroken is therefore lifted by the quantum deformation of the moduli
space.  It follows that if the Kahler potential is positive definite
and non-singular, supersymmetry is spontaneously
broken by the quantum deformation of the $SU(2)$ moduli space.

Supersymmetry breaking by the quantum deformation of the moduli space
can be illustrated by expanding about the enhanced symmetry point with
an unbroken $SU(2)^{\prime} \subset SU(3)$.  At this point, the
components of $\widehat{q}$ and $\widehat{\overline q}$ which satisfy
the constraint are eaten by the Higgs mechanism.  The low energy
theory is $SU(2)^{\prime}$ with doublets $u$ and $d$ and singlets
$S_u$ and $S_d$, all coming from the fields $\overline U$ and
$\overline D$, with $X_1=\Lambda _2^2S_u$ and $X_2=\Lambda _2^2 S_d$.
The scale $\Lambda _{2^\prime}$ of the low energy theory is related to
$\Lambda _3$ by the matching condition 
$\Lambda_{2^{\prime}}^5 = \Lambda_3^7 /
\Lambda_2^2$ and the superpotential \wsuiii\ yields the superpotential
of the low energy theory,
\eqn\ttlow{
W = {\Lambda_{2^{\prime}}^5 \over ud} + \lambda \Lambda_2^2 S_d.  }
The low energy theory also has $U(1)_R$ and $U(1)$ flavor symmetries.
The first term in \ttlow\ is interpreted as coming from an instanton
in the low energy $SU(2)'$ with one flavor.  This leads to a potential
which pushes $u$ and $d$ away from the enhanced symmetry point.
It follows from the
$U(1)_R$ and $U(1)$ flavor symmetries that the superpotential 
in the low energy theory only depends on
$S_d$ via the linear term in \ttlow.

Since $S_d$ is a canonically normalized field in this limit,
supersymmetry is broken by the $F$ component of $S_d$ with $V \sim |
\lambda^2 \Lambda_2 ^4|$.  The low energy theory \ttlow\ appears to
have a runaway direction labeled by $ud$, and pseudo-flat directions
labeled by $S_u$ and $S_d$.  In the effective theory, corrections to
the Kahler potential from the strong $SU(2)$ dynamics lift these
directions.  In addition, for sufficiently large expectation values,
the theory returns to a classical regime, in which the potential
coming from the $D$ and $F$ terms rises in all directions in field
space.  The natural scale for the expectation value along $ud$ is
therefore ${\cal O}(\Lambda_2)$, and for $S_u$ and $S_d$ either $0$ or
${\cal O}(\Lambda_2)$.  With these expectation values, in the limit
$\lambda \Lambda_2^2 \gg
\Lambda_{2^{\prime}}^5 / \Lambda_2^3$, which is equivalent to
$\lambda^{1/7} \Lambda_2 \gg \Lambda_3$, the $SU(2)^{\prime}$
dynamical superpotential in the low energy theory gives an
insignificant contribution to the vacuum energy, and does not plays a
role in the supersymmetry breaking.  In this limit, supersymmetry
breaking is associated with the quantum deformation of the $SU(2)$
moduli space, with the $SU(3)$ acting as a spectator.  Unfortunately,
because some of the fields are ${\cal O}(\Lambda_2)$, the Kahler
potential receives large incalculable corrections, and a quantitative
solution of the ground state of the effective theory in terms of the
$u$, $d$, $S_u$ and $S_d$ is not possible.

\newsec{$SU(N) \times SP(M)$ Generalizations}

Many chiral models with two gauge groups exhibit the behavior
illustrated 
in the previous section. 
In one limit, supersymmetry is broken
by a dynamically generated superpotential in one
gauge group while, in another limit, it is broken by the
quantum deformation of a moduli space by another gauge group.
The simplest such generalizations of the $SU(3) \times SU(2)$ model 
are given by 
theories with gauge group and matter content
\eqn\suspfields{
\matrix{  
\quad     & SU(N) \times SP({1 \over2}(N-1)) & \quad \cr
          &               &               \cr
P         & (N,N-1)       & \quad         \cr
L         & (1,N-1)       & \quad         \cr
\overline{Q}_i & (\overline{N},1)   & i=1,\dots,N-1, \cr 
}} with $N$ odd.  These theories have a classical moduli space of
vacua with the gauge group generically completely broken and light
moduli $X_i=PL\overline Q_i$ and $Z_{ij}=P^2\overline Q_i\overline Q_j
= - Z_{ji}$.  In addition, there is another gauge invariant, $Y = P^N
L$, which vanishes as a classical constraint.  The classical moduli
space degeneracy is lifted by the tree level superpotential
\eqn\wtreexz{
W_{tree}=\lambda X_1+\sum^{N-1}_{i,j>2}\gamma^{ij}Z_{ij},
}
where $\gamma^{ij} = - \gamma^{ji}$ has rank $N-3$.
This superpotential leaves invariant some global flavor symmetries
and a non-anomalous $U(1)_R$ symmetry. 
The form of 
the superpotential \wtreexz\ could be enforced by weakly gauging 
certain subgroups of the flavor symmetries. 
As discussed in \tools, the non-renormalizable terms 
in \wtreexz\ are
required for $N>3$ to completely lift the classical moduli space. 
All the $X_i$ and $Z_{1j}$ are lifted by the first term in 
\wtreexz\ while the remaining $Z_{ij}$ are lifted by the 
non-renormalizable terms. 
Classically, there is a supersymmetric ground state at the origin.

The superpotential in the quantum theory is 
\eqn\wnnmo{
W = {\Lambda_{SU}^{2N+1}\over \pf Z} + {\cal A} \left( Y -
\Lambda_{SP}^{N+1} \right) + W_{tree} ,} where ${\cal A}$ is a
Lagrange multiplier field.  The first term arises from an instanton in
the broken $SU(N)$ and the second enforces the quantum deformed
constraint $Y=P^N L= \Lambda _{SP}^{N+1}$ arising from the $SP(M)$
dynamics \intpou.

The non-renormalizable terms in \wtreexz\ introduce an additional
scale beyond the dynamically generated scales.  In order for
\suspfields\ to have a weakly coupled regime below the scale of the
non-renormalizable operators, we assume $\gamma^{ij} \ll
\Lambda_{SU}^{-1}$ and $\Lambda_{SP}^{-1}$.  The $Z_{ij}$ $i,j \geq 2$
moduli are lifted only by non-renormalizable terms and therefore have a
much smaller classical potential than the other moduli in this limit.
As a consequence, the quantum mechanical ground state develops large
expectation values along these directions.  For $\lambda \ll 1$ the
full theory can then be analyzed in terms of the resulting effective
theory.
With $Z_{1j}=0$, the expectation values $Z_{ij}$ 
have maximal rank $N-3$.  With this maximal rank, the gauge group is
classically broken to $SU(3) \times SU(2)$ with matter content given
by \ttfields\ and the ${1 \over 2}(N-2)(N-3)$ singlets $Z_{ij}$, $i,j
\geq 2$.  
The $2(N-3)$ fields coming from the components of $\overline{Q}_1$ and
$L$ which lie along the broken generators pair up and gain a large
Dirac mass from the renormalizable term in \wtreexz.  Along these
directions the theory therefore reduces to the $SU(3)
\times SU(2)$ model, with matter fields and a superpotential
corresponding precisely to the theory of \ads\ and the previous
section, along with the light singlets $Z_{i,j}$, $i,j\geq 2$, with
the superpotential term in \wtreexz.  The scales of the low energy
$SU(3)\times SU(2)$ theory are related to those of the original theory
by matching conditions $\widehat{\Lambda}_3^7 = \Lambda_{SU}^{2N+1} /
(\pfp Z)$, and $\widehat{\Lambda}_2^4 = \Lambda_{SP}^{N+1} /
\sqrt{\pfp Z}$, evaluated at the scale $(\pfp Z)^{1/(2N-6)}$,
where $\pfp$ is over the sub-matrix of $Z_{ij}$ with non-vanishing
rank $N-3$.  The dynamically generated superpotential \wnnmo\ agrees
with \wsuiii\ upon using these matching relations, rescaling the
Lagrange multiplier field as ${\cal A}^{\prime}={\cal A} \sqrt{\pfp
Z}$, and using $\pf Z = P^2 \overline{U}
\overline{D} (\pfp Z)$ and $Y = P^N L =\sqrt{\pfp Z} P^3 L$, where
$\overline{D}$ and $\overline{U}$ denote the components of
$\overline{Q}_1$ and a linear combination of the $\overline{Q}_i$ $i
\geq 2$ which lie in the directions of the $SU(3)$ generators.

As in the previous section, in the limit $\lambda \ll 1$ and
$\lambda^{1/7} \widehat{\Lambda}_2 \ll \widehat{\Lambda}_3$,
supersymmetry breaking in the effective theory is due to an instanton
in the broken $SU(3)$ and the elementary fields of the $SU(3)\times
SU(2)$ theory have $K\approx K_{cl}$.  In order to obtain the
dependence of the ground state expectation values and vacuum energy on
the dynamical scales and couplings, we note that in this limit the
superpotential is parametrically of the form
\eqn\WAB{ 
W \sim
{\Lambda_{SU}^{2N+1} \over A^4 B^{2N-6}} + \lambda A^3 + \gamma B^4 , 
}
where $A$ are the fields
\ttfields\ of the effective $SU(3) \times SU(2)$ theory while $B^4$
corresponds to the components of $Z_{ij}$ which gain large expectation
values.  Assuming $K\approx K_{cl} = A^\dagger A+B^\dagger B$, the
potential is then of the form
\eqn\VAB{ 
V \sim \big\vert {\Lambda_{SU}^{2N+1}
\over A^5 B^{2N-6}}+ \lambda A^2 \big\vert^2 + \big\vert
{\Lambda_{SU}^{2N+1} \over A^4 B^{2N-5}} + \gamma B^3 \big\vert^2, 
}
where the first term comes from $\partial_A W$ and the second from
$\partial_B W$.  The expectation values in the effective $SU(3)
\times SU(2)$ theory are determined by a balance within the
$\partial_A W$ term between the $SU(3)$ instanton and the
renormalizable superpotential term, which gives the scaling $A \sim
\widehat{\Lambda}_3 / \lambda^{1/7}
\sim (\lambda ^{-1}\Lambda_{SU}^{2N+1} B^{6-2N})^{1/7}$.
This leads to a potential for the $B$ fields of the form
\eqn\Bpot{
V \sim \big\vert {\lambda^5 \Lambda_{SU}^{4N+2} \over B^{4N-12}}
\big\vert^{2/7} + \big\vert \big({\lambda^4 \Lambda_{SU}^{3(2N+1)} 
\over B^{6N-11}}\big)^{1/7} + \gamma B^3  
\big\vert^2. 
}  
The first term is
just the contribution to the potential from the effective $SU(3)
\times SU(2)$ theory.  The remaining terms come from the $B$
dependence of the $SU(3)$ instanton and the non-renormalizable
superpotential.  Notice that the non-perturbative terms lift the
potential for small values of $B$, while the non-renormalizable terms
lift the potential at large expectation values.  The position of the
ground state is determined by a balance between these two types of
terms.  One possibility is that the $B$ dependence of the $SU(3)$
instanton and the non-renormalizable terms balance, with an
insignificant contribution from potential of the effective $SU(3)
\times SU(2)$ theory.  This occurs for $(\gamma \Lambda_{SU})^{2N+1}
\gg
\lambda^{2N+2}$.  In this limit the expectation values scale as $B
\sim (\lambda^4  (\gamma \Lambda_{SU})^{-7})^{1/(6N+10)} \Lambda_{SU}$,
giving a vacuum energy $V \sim |(\lambda^{12} (\gamma
\Lambda)^{6N-11})^{1/(3N+5)}
\Lambda_{SU}^4|$.
In order for the approximation $K = A^\dagger A+B^\dagger B$ to be
valid, the ground state must be weakly coupled and thus it should be
that $B \gg \Lambda_{SU}, \Lambda_{SP}$, which requires $(\gamma
\Lambda_{SU})^7 \ll \lambda^4$ and $(\lambda^4 (\gamma
\Lambda_{SU})^{-7})^{1/(6N+10)} \gg
\Lambda_{SP} / \Lambda_{SU}$.
In addition, the quantum modification of the moduli space in the
effective theory is unimportant for 
$\lambda^{1/7} \widehat{\Lambda}_2 \ll \widehat{\Lambda}_3$, which is
equivalent to $(\lambda^{-4}(\gamma \Lambda_{SU})^{(N-3)/(N+1)}
)^{1/(6N+10)}
\gg \Lambda_{SP} / \Lambda_{SU}$.  
Another simple situation to
consider is the opposite limit, $(\gamma \Lambda_{SU})^{2N+1} \ll
\lambda^{2N+2}$, where the ground state is determined by a balance
between the $B$ dependent potential of the effective $SU(3) \times SU(2)$
theory and the non-renormalizable terms, with an insignificant
contribution from the $B$ dependence of the $SU(3)$ instanton.  
In this case the expectation values scale as 
$B \sim (\lambda^5  (\gamma
\Lambda_{SU})^{-7} )^{1/(4N+9)} \Lambda_{SU}$, giving a vacuum energy $V
\sim | (\lambda^{15} (\gamma \Lambda_{SU})^{4N-12})^{2/(4N+9)}
\Lambda_{SU}^4 |$.  For a weakly coupled ground state, $B \gg
\Lambda_{SU}, \Lambda_{SP}$, which 
requires $(\gamma \Lambda_{SU})^7 \ll
\lambda^5$ and $(\lambda^5(\gamma \Lambda_{SU})^{-7})^{1/(4N+9)} \gg
\Lambda_{SP} / \Lambda_{SU}$.
Finally, the quantum modification of the moduli space in the effective
theory is unimportant in this limit for $(\lambda^{-3}(\gamma
\Lambda_{SU})^{(N-3)/(N+1)})^{1/(4N+9)}
\gg \Lambda_{SP} / \Lambda_{SU}$.

In the limit $\lambda^{1/7} \widehat{\Lambda}_2 \gg
\widehat{\Lambda}_3$, supersymmetry breaking in the effective theory
is due to the quantum deformation of the $SU(2)$ moduli space.  The
potential in the effective theory in this limit is $V \sim | \lambda^2
\widehat{\Lambda}_2^4 |$, giving a potential for the $B$ of the form
\eqn\VMB{ 
V \sim \vert { \lambda^2 \Lambda_{SP}^{N+1} \over B^{N-3}} \vert +
\vert \gamma B^3 \vert^2 .
}
A balance between these terms gives 
$B \sim (\lambda (\gamma \Lambda_{SP})^{-1})^{2/(N+3)} \Lambda_{SP}$, and
vacuum energy 
$ V \sim | (\lambda^6 (\gamma
\Lambda_{SP})^{N-3})^{2/(N+3)} \Lambda_{SP}^4 |$.  
In order for the
ground state to be weakly coupled, $B \gg \Lambda_{SP}, \Lambda_{SU}$,
which requires $(\lambda^{-1}  (\gamma \Lambda_{SP}))^{2/(N+3)} \ll 1,
\Lambda_{SU}/ \Lambda_{SP}$.

\newsec{Quantum Removal of Classical Flat Directions}

Most models of supersymmetry breaking are constructed to have a
$U(1)_R$ symmetry, and, as a result of $D$ and $F$ term constraints,
no classical flat directions.  The classical supersymmetric ground
state with unbroken $U(1)_R$ is then at the origin of field space.  If
a non-perturbative dynamical superpotential lifts the origin, the
$U(1)_R$ is spontaneously broken, and it follows that supersymmetry is
broken \refs{\ads, \nonren, \sn}.  
The examples of the previous two sections
break supersymmetry in this manner in the limit that the $SU(N)$
dynamics dominates.  Although the two conditions of having a $U(1)_R$
symmetry and no classical flat directions are generically sufficient
for dynamical supersymmetry breaking, they are not
necessary \refs{\nonren,\sn}.

In the present section we show that models which have a classical flat
direction can also break supersymmetry.  This is counter to the
usual intuition that quantum dynamics can not lift entire flat
directions but can, at best, lead to a runaway potential with a
supersymmetric ground state at infinity.  This more familiar situation
is bound to be the case when the entire gauge group is broken by the
scalar expectation values, leading to gauge dynamics with a strength
which tends to zero along the flat direction.  However, it is
possible for quantum dynamics to lift an entire flat direction,
without leading to a runaway vacuum, if part of the gauge group is not
broken along the flat direction.

Models which illustrate the situation discussed above are based on   
\eqn\susptfields{
\matrix{  
\quad     & SU(N) \times SP({1 \over2}(N+1)) & \quad \cr
          &                                  & \cr 
P         & (N,N+1)                          & \quad \cr 
L         & (1,N+1)                          & \quad \cr
\overline{Q}_i & (\overline{N},1)            & i=1,\dots,N+1 \cr 
}}
with $N$ odd.
There is a classical moduli space of vacua with the gauge group
generically completely broken.  The classical moduli 
are $X_i$ and $Z_{ij}$, 
as in the previous section, 
$\overline{B}^i = \epsilon^{i j_1 \dots j_N} 
\overline{Q}_{j_1} \cdots \overline{Q}_{j_N}$,
and $Y=P^NL$ (with $SU(N)$ indices contracted with an epsilon tensor)
subject to the classical relations $Y\overline B ^i=[XZ^{(N-1)/2}]^i$
(antisymmetrizing over the flavor indices of $X$ and $Z$) and $\pf
Z=0$.  These light fields correspond to the matter fields which are
left massless after the Higgs mechanism.  In addition, there are gauge
invariant fields $H_i=P^{N+1}\overline Q_i$, which classically satisfy
constraints $H_i=0$.

The tree level superpotential  
\eqn\wtreesp{
W_{tree}=\lambda X_1+\sum^{N+1}_{i,j>2}\gamma^{ij} Z_{ij},
}
leaves invariant some global flavor symmetries and a non-anomalous
$U(1)_R$ symmetry.
All the $X_i$, $Y$, and $Z_{1j}$ are lifted by the tree level term,
$Z_{ij}$ $i,j \geq 2$ are lifted by the non-renormalizable terms,
but the $\overline{B}^i$ are left undetermined by \wtreesp.
Classically there is moduli space of supersymmetric ground states
given by $X_i = Y = Z_{ij} =0$ and $\overline{B}^i$ arbitrary. 
These theories only satisfy one of the two conditions mentioned
above: there is a $U(1)_R$ symmetry but there are flat directions
which are unlifted classically.  
As shown below, in the quantum theory the tree level
superpotential \wtreesp\ actually does lift all the flat directions
and supersymmetry is broken.  This is possible because the $SP(M)$ gauge
group is unbroken along the entire classically unlifted  
flat direction.

The quantum theory is described by the above light fields with 
superpotential 
\eqn\suconfi{
W={1\over \Lambda _{SU}^{2N-1}} \left(\overline{B} H - \pf Z \right)
+{\cal A} \left( Y\pf Z-HXZ^{(N-1)/2}+\Lambda _{SU}^{2N-1}\Lambda
_{SP}^{N+4}\right) +W_{tree}, } where ${\cal A}$ is a Lagrange
multiplier field.  This form of the exact superpotential may be
obtained, much as in the examples of \ils, by considering the limit
$\Lambda _{SU}\gg \Lambda _{SP}$.  In this limit, the $SP(M)$ is
weakly gauged at the scale $\Lambda_{SU}$.  The $SU(N)$ theory
therefore has $N_f = N+1$ flavors of fundamentals and confines into a
theory of ``mesons'' $M_i=P\overline Q_i$ which transform as $SP(M)$
fundamentals (as always, we display the flavor index, suppressing the
$SP(M)$ gauge index), ``baryons'' $B=P^N$ which transform as an
$SP(M)$ fundamental, and ``antibaryons'' $\overline{B}^i$ which are
$SP(M)$ singlets.  It follows from \nati\ that these 
confined fields have a
superpotential corresponding to the first term in \suconfi.  Below the
scale $\Lambda_{SU}$, the $SP(M)$ has $N_f = M+1$ flavors ($2M+2=N+3$
matter fields) in the fundamental representation, corresponding to the
fields $\widehat{M}_i = M_i / \Lambda_{SU}$, the field $\widehat{B} =
B / \Lambda_{SU}^{N-1}$, and the field $L$.  Notice that, because of
the extra confined degrees of freedom which becomes massless at the
origin, the $SP(M)$ theory below the scale $\Lambda_{SU}$ has one more
flavor than the underlying high energy theory.  The scale $\widehat
\Lambda _{SP}$ of the low energy theory with one extra flavor is
related to $\Lambda _{SP}$ by the matching relation $\Lambda
_{SP}^{N+4}=\Lambda _{SU} \widehat 
\Lambda _{SP}
^{N+3}$ at the confining scale $\Lambda _{SU}$.  $SP(M)$ with $N_f =
M+1$ flavors has a quantum deformed moduli space \intpou, which is
enforced by the second term in \suconfi.  The scale $\Lambda_{SU}$
appears in the quantum deformed constraint because factors of $\Lambda
_{SU}$ in $\widehat \Lambda_{SP}$, $\widehat M_i$, and $\widehat B$,
reflecting the matching conditions and the fact that some of the
canonically normalized $SP(M)$ flavors are composites 
associated with the $SU(N)$ confinement.

We now consider the limit in which the $SP(M)$ dynamics dominate,
$\Lambda_{SU} \ll \Lambda_{SP}$.  The first term in \suconfi, arising
{}from the $SU(N)$ confinement, becomes large in this limit.  Imposing
equations of motion which receive contributions from this term amounts
to enforcing classical constraints on the $SU(N)$ moduli space.  In
particular, $\overline{B}^i$ and $H_i$ get a large mass in this limit
and hence can be integrated out by imposing their equations of motion,
leading to the classical constraint $H_i=0$ along with
$\overline{B}^i=\Lambda _{SU}^{2N-1}{\cal A}(XZ^{(N-1)/2})^i$.  In
addition, the Pfaffian of the $Z_{ij}$ equations of motion, and
$H_i=0$, imply ${\cal A} = 1/(\Lambda_{SU}^{2N+1} Y)$ in this limit.
Together with the $H_i$ equations of motion this yields the classical
constraint $Y\overline B ^i=[XZ^{(N-1)/2}]^i$, and the superpotential
\suconfi\ becomes
\eqn\wspinst
{ W={\Lambda _{SP}^{N+4}\over Y} + {\cal A}^{\prime} \left( Y ~\pf Z +
\Lambda_{SU}^{2N-1} \Lambda_{SP}^{N+4} \right) +W_{tree}, } where
${\cal A}^{\prime}$ is a different Lagrange multiplier field.  The
first term is generated by an instanton in the broken $SP(M)$.  The
second term enforces the quantum deformed constraint $Y~\pf Z = -
\Lambda_{SU}^{2N-1} \Lambda_{SP}^{N+4}$ associated with the $SP(M)$
dynamics.  In order for the $SP(M)$ instanton superpotential to be
finite requires $Y \neq 0$.  For $\Lambda_{SU} =0$, the constraint
enforced by the ${\cal A}^{\prime}$ equation of motion therefore
yields the classical constraint $\pf Z=0$ associated with the
classical $SP(M)$ moduli space,\foot{The classical constraint $\pf
Z=0$ can also be obtained for $\Lambda_{SU} \neq 0$ in the theory with
$W_{tree}=0$ by integrating out $Y$ and enforcing the ${\cal
A}^{\prime}$ equation of motion.  This gives the superpotential $W= -
\pf Z / \Lambda_{SU}^{2N-1}$.  The $Z_{ij}$ equations of motion yield
the classical constraint $\pf Z=0$, and the ${\cal A}^{\prime}$
equation of motion implies $Y \rightarrow \infty$.}  and the
superpotential becomes
\eqn\wspicl{
W = {\Lambda_{SP}^{N+4} \over Y} + W_{tree}.
}
The dynamically generated instanton superpotential is just that which
would arise over the classical moduli space for $\Lambda_{SU}=0$.

The lifting of the classical flat directions $\overline{B}^i$
and supersymmetry breaking can be seen in the 
$\Lambda_{SU} \ll \Lambda_{SP}$ limit in the effective theory
\wspicl\ subject to the classical constraints among the 
gauge invariants.  In this limit the $SP(M)$ instanton potential in
\wspicl\ leads to $Y \neq 0$ in the quantum theory.  For $Y\neq 0$,
the constraint $Y\overline B=XZ^{(N-1)/2}$ can be used to relate the
$\overline{B}^i$ flat directions to $X_i$ and $Z_{ij}$ flat
directions.  Because $W_{tree}$ lifts the $X_i$ and $Z_{ij}$ flat
directions, it eliminates all flat directions in the quantum theory.
Since $Y$ is also lifted by $W_{tree}$, there is a stable ground state
with the $U(1)_R$ symmetry is spontaneously broken.  It follows that
supersymmetry is broken.  In this limit the classical flat directions
are lifted by the tree level superpotential in the presence of the
dynamically generated $SP(M)$ superpotential.  We conclude that it is
possible for theories with classical flat directions to dynamically
break supersymmetry with a stable vacuum.
 
\newsec{Resolving the Origin}

In order to analyze the issue of supersymmetry breaking, it is crucial
that all of the relevant light fields entering in the low energy
effective theory be properly identified.  Neglecting some light fields
on a moduli space can lead to the mistaken conclusion that a theory
breaks supersymmetry while, upon proper inclusion of all relevant
fields, it is seen that the theory actually has a supersymmetric
vacuum.  As a simple example, consider a Wess-Zumino theory with
chiral superfields $L$ and $H$ with canonical Kahler potential
and superpotential $W=LH^2$.  The field $L$ is massless and can have
arbitrary expectation value.  For $L\neq 0$, $H$ is massive and can be
integrated out, yielding the low energy superpotential $W=0$.  Adding
a superpotential $W=\mu^2 L$ to the low energy theory with $H$
integrated out appears to break supersymmetry, $\partial _LW\neq 0$.
But in the theory with $H$ included, there is a supersymmetric ground
state at $L=0$, and $H^2=-\mu^2$.  In order to realize that
supersymmetry is unbroken, a low energy observer must know that, in
addition to the light field $L$, there is another field $H$
which becomes massless at $L=0$.\foot{Exactly this 
situation occurs for $N=2$ Yang-Mills theory broken explicitly to
$N=1$ Yang-Mills theory by giving a mass to the adjoint $\Phi$.  There
$L\sim \Tr~ \Phi ^2$ and $H$ corresponds to the monopoles which are
massless at strong coupling \swi.}

Models which break supersymmetry by a dynamically generated
superpotential become strongly coupled for small field values since
the dynamical superpotential becomes singular at the origin.  A
quantitative analysis does not seem possible in this region.  One
might worry that such models could develop another branch in this
strong coupling region along which supersymmetry is unbroken, as in
the toy example given above.  One way to gain confidence that there
are no such strong coupling subtleties, which could lead to
supersymmetric vacua near the origin, is to consider theories with
additional, massive, vector-like matter integrated in.  Additional
matter makes the gauge dynamics more weakly coupled, leading to a more
weakly coupled description of the supersymmetry breaking.  The
dynamics of the original theory is recovered in the limit that the
mass of the vector matter is taken much larger than all dynamical
scales in the theory.
Extensions of known supersymmetry breaking models with additional
vector-like matter have been considered previously
in \refs{\pt, \hitoshi, \poustr}.

Depending on how many massive vector-like matter fields are added, it
is possible to realize supersymmetry breaking in the low energy theory
via a variety of mechanisms.  In the present context it is most useful
to add enough vector-like matter so that all the gauge groups become
confining.  The region of field space which was strongly coupled in
the original model is then weakly coupled in the theory with
additional vector matter.  Extra confined degrees of freedom do in
fact become light near the origin.  However, since the gauge groups
are confined, if supersymmetry is broken it must be realized as a tree
level effect for the confined fields.  The low energy theory with
additional vector like matter then amounts to an O'Raifeartaigh-type
model, in which supersymmetry breaking can be easily verified.

As an example, consider the $SU(3)\times SU(2)$ theory with 
matter content
\eqn\ttexfields{
\matrix{  
\quad   & SU(3) \times SU(2) \qquad & \cr
        &             \qquad & \cr
P       & (3,2)       \qquad & \cr
L_a       & (1,2)       \qquad &a=1,\dots, 3\cr
Q         & (3,1)       \qquad &\cr
\overline{Q}_i & (\overline{3},1)\qquad &i=1,\dots, 3.
\cr
}} 
This is just the $SU(3) \times SU(2)$ model of 
\ads\ and section 3 with an extra,
vector-like, $SU(3)$ flavor and an extra, vector-like, $SU(2)$ flavor.
The classical moduli space is parameterized by the invariants $M=P^2Q$,
$X_{a,i}=PL_a\overline Q_{i} $, $Y_a=P^3L_a$, $M_i=Q\overline Q_i$,
$B=P^3L_aL_bL_c\epsilon ^{abc}$, 
$B^a=Q(PL_b)(PL_c)\epsilon ^{abc}$, 
$\overline B= \overline Q_i\overline Q_j\overline Q_k\epsilon ^{ijk}$, 
$\overline B^i=\epsilon^{ijk}P^2\overline Q_j\overline Q_k$, 
and $V^a=\half \epsilon ^{abc}L_bL_c$
subject to the classical relations $M \overline{B} = - M_i
\overline{B}^i$, $Y_a \overline{B} = - X_{a,i} \overline{B}^i$, $M B =
- Y_a B^a$, $M_i B = - X_{a,i} B^a$, $B \overline{B} = \det X$, $B^a
\overline{B} = -\half M_i X_{b,j} X_{c,k} \epsilon^{abc} \epsilon^{ijk}$,
$B \overline{B}^i = -\half Y_a X_{b,j} X_{c,k} \epsilon^{abc}
\epsilon^{ijk}$, and $B^a \overline{B}^i = \half M X_{b,j} X_{c,k}
\epsilon^{abc} \epsilon^{ijk} -  Y_b M_j X_{c,k} \epsilon^{abc}
\epsilon^{ijk}$.  The tree level superpotential
\eqn\wtreemetc{
W_{tree}=\lambda X_{1,1}+m_QM_3+m_LV^1
}
leaves invariant non-anomalous $U(1)_R$ and $U(1)$ flavor
symmetries, and completely lifts the moduli space. 
Classically there is a supersymmetric ground state at the origin.

The quantum theory is described by the above light fields 
with superpotential
\eqn\sufconf{\eqalign{
W 
=& -{1\over \Lambda _2^3}(B-Y_aV^a) 
+ {1\over \Lambda _2^3\Lambda _3^6}\big(MB\overline B
+ Y_aB^a\overline B+M_iB\overline B^i
+ X_{a,i}B^a\overline B^i \cr
& - M\det X
 +\half Y_aM_iX_{b,j}X_{c,k}\epsilon ^{abc}\epsilon ^{ijk} \big)
+ W_{tree}.
\cr}}
This form of the exact superpotential may be obtained by considering
the limit $\Lambda_2 \gg \Lambda_3 \gg m_Q , m_L$.
In this limit, the $SU(3)$ is weakly coupled at the scale 
$\Lambda_2$.
The $SU(2)$ theory therefore has 3 flavors of fundamentals
and confines into a theory of ``mesons'' 
$V^a$ which are $SU(3)$ singlets, 
$PL/\Lambda _2$ in the ${\bf 3}$ of $SU(3)$, and $P^2/\Lambda _2$ in the 
$\overline{\bf 3}$ of $SU(3)$.
The superpotential for these confined fields corresponds
the the first term in \sufconf\ \nati.
Because of the extra confined $SU(2)$ mesons which become
massless at the origin, the $SU(3)$ theory below the scale
$\Lambda_2$ has one more flavor than the underlying high energy
theory. 
The scale $\widehat{\Lambda}_3$ of the low energy theory
with one extra flavor is related to $\Lambda_3$ by the
matching condition $\Lambda_3^6 = \Lambda_2 \widehat{\Lambda}^5_3$
at the confining scale $\Lambda_2$.
The $SU(3)$ theory with 4 flavors confines \nati\ into a theory of 
``mesons'' 
$\widehat{M} = M  / ( \Lambda_2 \Lambda_3)$,
$\widehat{X}_{a,i} = X_{a,i} / (\Lambda_2 \Lambda_3)$,
$\widehat{Y}_a = Y_a / (\Lambda_2^2 \Lambda_3)$,
$\widehat{M}_i = M_i / \Lambda_3$,
and ``baryons'' 
$\widehat{B} = B / (\Lambda_2^3 \Lambda_3^2)$,
$\widehat{B}^a = B^a / (\Lambda_2^2 \Lambda_3^2)$,
$\widehat{\overline{B}} = \overline{B} / \Lambda_3^2$,
and 
$\widehat{\overline{B}}^i = \overline{B}^i / (\Lambda_2 \Lambda_3^2)$.
These fields, along with the $SU(2)$ meson 
$\widehat{V}^a = V^a / \Lambda_2$, make up the canonically normalized
confined fields which are light near the origin. 
The $SU(3)$ confining superpotential is given by the second
term in \sufconf.
As in the previous section, the scale $\Lambda_2$ arises
in this term because of the matching
relation and normalization factors of $\Lambda _2$ 
coming from the $SU(3)$ flavors which are composites
associated with the $SU(2)$ confinement.

For $m_Q, m_L \ll \Lambda_3, \Lambda_2$, and $\lambda \ll 1$ the
expectation values in the ground state are much smaller than the
dynamical scales, and the superpotential \sufconf\ is the relevant one
to consider.  In this limit the theory just amounts to an
O'Raifeartaigh-type model for the confined degrees of freedom, with
tree level couplings \sufconf.  Supersymmetry is broken if all the
auxiliary $F$ components can not simultaneously vanish.  Some of the
$F$ terms are \eqn\fterms{ \eqalign{(X^{a,i})^*_F & ={1\over \Lambda
_3^6\Lambda _2^3} (B^a\overline B^i -\half MX_{b,j}X_{c,k}\epsilon
^{abc}\epsilon ^{ijk}+ Y_bM_jX_{c,k}\epsilon ^{abc}\epsilon
^{ijk})-\lambda\delta ^{a,1}\delta ^{i,1}\cr (M^i)^*_F & ={1\over
\Lambda _3^6\Lambda _2^3}(B\overline B^i+\half Y_aX_{b,j}X_{c,k}\epsilon
^{ijk}\epsilon ^{abc})+m_Q\delta ^{i,3}\cr (V_a)^*_F&={1\over \Lambda
_2^3}Y_a+m_L\delta _{a,1} \cr (\overline B)^*_F&={1\over \Lambda
_2^3\Lambda _3^6}(MB+Y_aB^a).\cr }} Note first that
$Y_a(X^{a,3})^*_F+M(M^3)^*_F-\overline B^3(\overline B)^*_F=Mm_Q$.
Vanishing $F$ terms therefore require $M=0$ for $m_Q\neq 0$.  It then
follows that $B^a(V_a)^*_F-\Lambda _3^6(\overline{B})^*_F=m_LB^1$ and,
therefore, $B^1=0$ for $m_L\neq 0$.  In addition, $(V_a)^*_F =0$
implies $Y_a=0$, $a \neq 1$.  Finally, it follows from $M=B^1=Y_a=0$,
$a \neq 1$, that $(X^{1,1})^*_F=-\lambda$. So all the $F$ terms in
\fterms\ can not vanish for non-zero $m_Q$, $m_L$, and $\lambda$, and
supersymmetry is therefore broken.  In any theory of confined fields
which breaks supersymmetry,
there must be at least one term in the superpotential
which is linear in a confined field.  Otherwise, all fields can sit at
the origin with supersymmetry unbroken.

For $m_Q, m_L \gg \Lambda_2, \Lambda_3$ 
the additional vector-like matter can be
integrated out, yielding the $SU(3)\times SU(2)$ model \ttfields\
with superpotential \wsuiii.
This can be verified explicitly in this limit 
by using the matching relations
$\widehat{\Lambda}_3^{{\prime} 7} = m_Q \Lambda_3^6$
and $\widehat{\Lambda}_2^{{\prime} 4} = m_L \Lambda_2^3$, 
and integrating out all the fields in \sufconf\ but 
$Z = \overline{B}^3$, $X_1 = X_{1,1}$, $X_2=X_{1,2}$, and
$Y=Y_1$, which are just the standard $SU(3)\times SU(2)$
invariants.
We therefore see explicitly that supersymmetry is broken 
in both the large and small mass limit. 
Any additional branches at strong coupling 
in the original $SU(3) \times SU(2)$ model
along which supersymmetry
could be restored would presumably survive 
in the theory with additional vector matter, and appear at the origin
as additional massless fields 
in the weakly coupled, small mass, limit.
This is not the case, and supersymmetry is indeed broken 
at tree level in this weakly coupled limit.
This gives confidence that there are no such strong coupling
subtleties in the $SU(3) \times SU(2)$ model. 
Extension to other models which break supersymmetry by 
 dynamically generated superpotentials are straightforward. 

\newsec{Conclusions}

The quantum deformation of a classical moduli space 
can lead to supersymmetry breaking. 
This occurs if the quantum deformed constraint associated
with the moduli space is inconsistent with a stationary
superpotential.  The vacuum energy and 
auxiliary components for some fields then do not vanish. 
It may at first sight seem surprising that supersymmetry
can be broken even though the dynamical superpotential on the classical
moduli space exactly vanishes. 
In fact, the absence of a dynamical superpotential 
over the classical moduli space has often been 
given as a ``proof'' that supersymmetry could not be broken
in such theories. 
However, it is possible that the tree level auxiliary components
only vanish at points on the classical moduli space which 
are removed by the quantum deformation, thus leading
to supersymmetry breaking.
In some instances it is possible to analyze models which 
break supersymmetry by this mechanism by expanding 
about enhanced symmetry points on the quantum moduli space. 
The Kahler potential for the non-perturbative fields
which become light at these points is approximately
canonical, and often allows a quantitative description of
the relevant degrees of freedom in the ground state.

The $SU(2)$ model discussed in section 2 has a number
of interesting features aside from demonstrating supersymmetry
breaking
by the quantum modification of a moduli space.
It is the first example of a renormalizable model in which 
singlet fields participate directly in the supersymmetry 
breaking. 
This can have phenomenological applications to both
hidden sector and gauge mediated supersymmetry breaking.
Hidden sector singlets which participate directly in 
supersymmetry breaking 
are required in renormalizable hidden sector models 
in order to obtain dimension three soft terms of order
the weak scale \refs{\bkn}.
Without this, gauginos are much lighter than
the weak scale in such scenarios \refs{\bkn,\smallmu}.
Likewise,
in gauge mediated supersymmetry breaking, singlets with
both scalar and $F$ components of order
$S \sim \sqrt{F_S}$ are required in many schemes 
\refs{\tools,\visible}.
The singlets couple 
through Yukawa couplings, $hSQ _m \overline{Q}_m$, 
to vector quarks $Q_m$ and $\overline{Q}_m$ which transform under the
standard model gauge group.
The vector quarks act as messengers of supersymmetry breaking;
integrating them out generates soft and explicit supersymmetry
breaking in the low energy theory. 
In order to obtain $S \sim \sqrt{F_S}$
previous schemes have invoked a secondary messenger sector
with nonzero $D$ terms to couple 
the supersymmetry breaking sector with the 
singlets \refs{\tools,\visible}.
In the $SU(2)$ model given here, however,
it is possible that the singlets have a minimum with 
$S \sim \sqrt{F_S} \sim \Lambda_2$
for $\lambda \sim 1$.  Again, the location of the minimum along the
pseudo-flat direction depends on
the quantum corrections to the Kahler potential which,
unfortunately, 
can not be computed quantitatively.
It is a natural possibility, though, for a minimum to exist at 
${\cal O}(\Lambda_2)$. Note, however, that simply adding the messenger
quark coupling $hSQ _m\overline{Q}_m$ to the model of section 2 would
introduce a supersymmetric ground state with $F_S=0$ and 
$Q_m\overline{Q_m} \neq 0$.
 The desired minimum with $F_S \neq 0$ and $Q_m\overline{Q}_m =0$ 
could only be a local minimum.  In any case, this
model demonstrates that it is possible in principle to consider
schemes without the secondary messenger sector.

Another interesting feature of the $SU(2)$ model is that it is
non-chiral with respect to the gauge group, an aspect stressed in
\talk\ and very recently also in \ref\scoo{K. Izawa and T. Yanagida,
hep-th/9602180.}.  This appears to conflict with the Witten index
\index\ argument that ${\rm Tr}(-1)^F \neq 0$ in non-chiral models and
supersymmetry is, therefore, unbroken.  It is, in fact, possible to
have ${\rm Tr}(-1)^F=0$ in non-chiral models when, as in the $SU(2)$
model, there is a pseudo-flat direction which extends to infinity.
This allows the index to change under small deformations of the model
since vacua can move in or out from infinity along this direction.
When the pseudo-flat direction degeneracy is lifted by quantum effects
as discussed in section 2, the vacuum has ${\rm Tr}(-1)^F=0$ and
supersymmetry is broken.  If, however, the singlets $S^{ij}$ are given
a mass $\epsilon$ by adding a term $\epsilon~ \pf S$ to the
superpotential, the index becomes ${\rm Tr}(-1)^F=2$ as expected for
$SU(2)$ \index, and there are two supersymmetric vacua.  To understand
this change in the index, consider the low energy theory along the
pseudo-flat direction $S_0$ with $W=\pm 2\lambda \Lambda
_2^2S_0+\epsilon S_0^2$.  For $\epsilon=0$, the vacuum energy along the
entire $S_0$ direction is $V\sim |\lambda ^2\Lambda _2^4 |$.  However,
for $\epsilon \neq 0$, there are two supersymmetric ground states at
$S_0=\pm \lambda\Lambda _2^2/\epsilon$.  For $\epsilon \rightarrow 0$
these ground states are sent to $\infty$ along the flat direction.  In
this way, ${\rm Tr}(-1)^F$ is discontinuous at $\epsilon=0$; the
theory with $\epsilon=0$ (enforced by discrete symmetries or by
gauging certain subgroups of the flavor symmetry) breaks supersymmetry
whereas that with $\epsilon\neq 0$ does not.  This pathologic behavior
of the index occurs only when there is a pseudo-flat direction in the
quantum theory and does not occur for the chiral models discussed in
this paper.  Such behavior is expected for any non-chiral model of
dynamical supersymmetry breaking.  For example, pure gaugino
condensation with a moduli dependent gauge kinetic function exhibits
the same behavior.

The models of section 5 provide examples in which a flat direction
remains in the classical theory, but is lifted, with supersymmetry
broken, in the quantum theory.  The standard list of ``requirements''
for stable dynamical supersymmetry breaking was assumed to include:
chiral matter, no classical flat directions which extend to infinity,
a dynamically generated superpotential, and a spontaneously broken
$U(1)_R$ symmetry.  The difficulty in finding models which
simultaneously satisfy all these requirements lead to the belief that
dynamical supersymmetry breaking was difficult and perhaps not
generic.  As shown in Refs. 
\refs{\nonren, \sn} a $U(1)_R$ symmetry is in fact not
necessary.  We now see that the other ``requirements'' are also not
necessary.  It would be interesting to find a single model which
satisfies none of the standard ``requirements'' but, nevertheless, breaks
supersymmetry.

Models which break supersymmetry by a dynamically generated 
superpotential become
strongly coupled near the origin.
As discussed in section 6,
integrating in extra vector-like matter makes the gauge dynamics
more weakly coupled and allows the region of strong coupling
to be resolved.  
New vacua for which supersymmetry
might be restored do not appear in the theory with
vector-like matter integrated in. 
It is in fact unlikely that such additional massless states could appear
as functions of the known moduli without ruining the 
't Hooft anomaly matchings. 
Although this is of course not a proof, it does provide confidence
that there are no such branches at strong coupling in the 
original theory without vector matter. 

Finally, we note that although the discussion throughout has
been in terms of rigid supersymmetry in which supersymmetry
breaking implies a non-zero vacuum energy, generalization
of any of the mechanisms discussed here to supergravity is
straightforward.

\centerline{{\bf Acknowledgments}}

We would like to thank M. Dine, A. Nelson, R. Rattazzi, and N. Seiberg
for useful discussions.  The work of K.I. was supported by NSF grant
PHY-9513835 and the W.M.  Keck Foundation.  The work of S.T. was
supported by the Department of Energy under contract DE-AC03-76SF00515
and the National Science Foundation under grant PHY-94-07194.  We
would also like to thank the Aspen Center for Physics 
and Institute for Theoretical Physics where this work
was partially completed.

\listrefs
\bye